\documentclass[superscriptaddress,twocolumn,prl,showpacs]{revtex4}
\usepackage[dvips]{graphicx}
\usepackage{epsfig}
\usepackage{psfrag}
\usepackage{amsmath,amsfonts,amssymb,bm,ulem}

\usepackage[usenames]{color}

\begin{document}
\title{Stability of Bose Einstein condensates of hot magnons in YIG}
\author{I. S. Tupitsyn}
\affiliation{Pacific Institute of Theoretical Physics, University of
British Columbia, 6224 Agricultural Road, Vancouver, BC Canada, V6T 1Z1}
\author{P. C. E. Stamp}
\affiliation{Pacific Institute of Theoretical Physics, University of
British Columbia, 6224 Agricultural Road, Vancouver, BC Canada, V6T 1Z1}
\author{A. L. Burin}
\affiliation{Department of Chemistry, Tulane University, New
Orleans, LA 70118, USA}

\date{\today}

\begin{abstract}
We investigate the stability of the recently discovered room
temperature Bose-Einstein condensate (BEC) of magnons in Ytrrium
Iron Garnet (YIG) films. We show that magnon-magnon interactions
depend strongly on the external field orientation, and that the BEC
in current experiments is actually metastable - it only survives
because of finite size effects, and because the BEC density is very
low. On the other hand a strong field applied perpendicular to the
sample plane leads to a repulsive magnon-magnon interaction; we
predict that a high-density magnon BEC can then be formed in this
perpendicular field geometry.

\end{abstract}

\maketitle

In a remarkable very recent discovery \cite{demo06}, a {\it Room
Temperature} Bose-Einstein condensate (BEC) of magnon excitations
was stabilized for a period of roughly $1 \; \mu s$ in a thin slab
of the well-known insulating magnet Yttrium Iron Garnet (YIG). The
density of magnons was quite low ($n \sim 10^{-4}$ per lattice
site), and the density $n_o$ of the BEC was apparently unknown, but
$n_o/n \ll 1$. This result can be understood naively in terms of a
weakly-interacting dilute gas of bosons, provided that (i) one
assumes that the number of magnons is conserved, so their chemical
potential may be non-zero, and (ii) the interactions between them
are repulsive (attractive interactions favour depletion of the BEC,
causing a negative compressibility and instability of the BEC
\cite{BEC}). In the experiment, the magnon dispersion was controlled
both by the sample geometry and external magnetic field, in such a
way that magnon-magnon collisions conserved magnon number - the
decay of the BEC was then attributed to spin-phonon couplings
\cite{demo06}. The magnons in the experiment had rather long
wavelengths, of order a $\mu$m - hitherto such excitations have been
treated entirely classically \cite{walker}.

This experiment raises a number of important questions, not least of
which concern the kind of superfluid properties possessed by a BEC
of such unusual objects. However a much more basic question about
the stability of the system must first be answered. In fact we find
the rather startling result that under the conditions of the
experiments reported so far, these interactions were actually {\it
attractive}, ie., the BEC ought to be unstable! We shall see that
because of the particular geometry used, the BEC is actually
metastable to thermally activated or tunneling decay, and that it
only survives because its density is low - above a critical density,
given below, it is absolutely unstable. However we also find that by
changing the field configuration in the system one can make the
interactions repulsive, and the BEC should then stabilize at a much
higher density - opening the way to much more interesting
experiments.

YIG is one of the best characterized of all insulating magnets
\cite{YIG}. It is cubic, with lattice constant $a_o = 12.376 {\AA}$,
ordering ferrimagnetically below $T_c = 560K$. At room temperature
the long-wavelength properties can be understood using a Hamiltonian
with ferromagnetic exchange interactions between effective 'block
spins' ${\bf S_j}$, one per unit cell, whose magnitude $S_j = \vert
{\bf S_j} \vert = a_o^3 M_s/\gamma$, with $\gamma = g_e \mu_B$, is
defined by the experimental saturated magnetisation density $M_s$
($M_s \approx 140 \; G$ at room temperature; with $g_e \approx 2$,
one has $S_j \approx 14.3$), along with dipolar couplings between
these; the resulting lattice Hamiltonian takes the form:
\begin{eqnarray}
\widehat{H} &=& - \gamma \sum_{i}{\bf S}_{i} \cdot {\bf H}_o - J_o
\sum_{i,\delta} {\bf S}_i {\bf S}_{i+\delta}
\nonumber \\
&+& U_d \sum_{i \neq j} \frac{{\bf S}_i  {\bf S}_j - 3 ({\bf S}_i
\cdot {\bf n}_{ij}) ({\bf S}_j \cdot {\bf n}_{ij})} {|r_{ij}|^3},
 \label{H_bare}
\end{eqnarray}
where the sums $i$, $j$ are taken over lattice sites at positions
${\bf R}_i$, etc., {\bf $\delta$} denotes nearest-neighbor spins,
${\bf r}_{ij} = ({\bf R}_i - {\bf R}_j)/a_o$, and ${\bf n}_{ij} =
{\bf r}_{ij}/\vert {\bf r}_{ij} \vert$. The nearest-neighbour
dipolar interaction $U_d = \gamma^2 / a_o^3 \approx 1.3 \times 10^{-3} K$.
The isotropic exchange $J_o$ is determined experimentally from $J =
2 S J_o a^2_o \approx 0.83 \times 10^{-28} \; erg \; cm^2$ at room
temperature. One then has $J_o \approx 1.37 \; K$, and $U_d / J_o
\approx 0.95 \times 10^{-3}$.

In what follows we set up a theoretical description of the BEC,
taking into account the external field, dipolar and exchange
interactions, and boundary conditions in the finite geometry. We
evaluate the interactions and the BEC stability for 2 different
field configurations; the general picture then becomes clear.

\vspace{1mm}

\begin{figure}[h]
\vspace{-2.7cm}
\includegraphics[width=8cm]{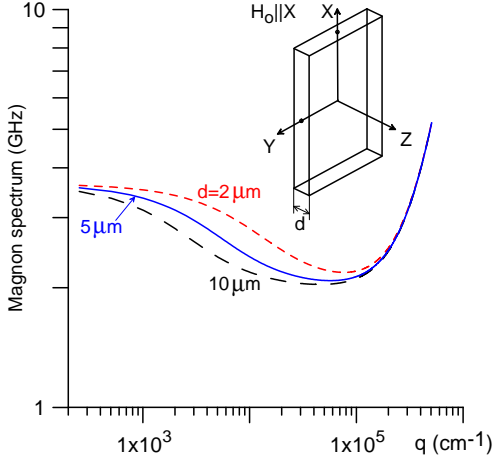}
\vspace{-2.8cm} \caption{ (color online). The magnon spectrum,
Eq.\ref{home_npb}, for $d = 2, \; 5$ and $10 \; \mu m$ at $H_o = 700 \; Gauss$. 
The inset shows the sample geometry.}
\label{Fig1}
\end{figure}

{\bf (i) In-plane field}: All experiments so far have had ${\bf
H_o}$ in the slab plane. The combination of exchange, dipolar, and
Zeeman couplings then leads to a magnon spectrum $\omega_{\bf q}$
shown in Fig.\ref{Fig1}, in which the competition between dipolar
and exchange interactions leads to a finite-$q$ minimum in
$\omega_{\bf q}$ at a wave-vector $Q \sim 1/d$, where $d$ is the
slab thickness. To completely specify $\omega_{\bf q}$ and the
inter-magnon interactions one needs boundary conditions, which can
involve partial pinning of the surface spins
\cite{rado59,patton,KALSLAV}. Demokritov et al \cite{demo06} assume
free surface spins, implying that (i) $\vert (\partial M({\bf
r})/\partial {\bf r})\cdot {\bf n}_s \vert = 0$ when ${\bf r}$ is at
the surface; here ${\bf n}_s$ is the normal to the surface, and (ii)
that the allowed momenta along $\hat{z}$ (see Fig.\ref{Fig1}, inset)
are $q_{\perp} = n_{\perp} \pi / d$, leading to different magnon
branches labelled by $n_{\perp}$. For now we assume a continuous
in-plane momentum, and later discuss the effect of in-plane
quantization; and we assume $n_{\perp} = 0$, taking the
lowest-energy magnon branch.

The above assumptions yield an $\omega_{\bf q}$ which agrees with
experiment \cite{demo06}, and which for ${\bf H}_o \parallel \hat{x}$ takes
the form \cite{KALSLAV}
\begin{equation}
\hbar \omega_q = \left[ (\gamma H_i + J q^2) (\gamma H_i + J q^2 + \hbar
\omega_M {\cal F}_q) \right]^{1/2},
\label{home_npb}
\end{equation}
here $H_i = H_o - 4 \pi N_x M_s$ is the internal field, $N_x$ the
demagnetization factor (for a slab in the $xy$-plane, $N_x=N_y=0$,
$N_z=1$), $\hbar \omega_M = 4 \pi \gamma M_s$ (for YIG, $\hbar \omega_M
\approx 0.236 \; K$ at room temperature), and $\hbar \omega_H = \gamma H_i$.
The form of the dimensionless function ${\cal F}_q$ depends on the
direction of ${\bf q}$. In the important case when ${\bf q}$ is parallel
to $H_o$, ie., along $\hat{x}$, one has
\begin{equation}
{\cal F}_q  \rightarrow \left( 1 - e^{- q_{x} d} \right) / (q_{x} d),
\label{F_oo}
\end{equation}
In the experiment \cite{demo06} magnons are argued to condense at
the minima $q_x = Q$; when $d = 5 \; \mu m$, and $H_o = 700 \; Gauss$, 
one has $|Q| \approx 5.5 \times 10^4 \; cm^{-1}$.

We now set up a theoretical description of the BEC, including all
4-magnon scattering processes (3-magnon scattering is excluded by
the kinematics), using a generalized Bogoliubov quasi-average
technique \cite{Belyaev58} to incorporate the BEC. Defining magnon
operators $b_{\bf q}, b_{\bf q}^{\dagger}$, a magnon BEC at $q=Q$,
with $N_o$ condensed magnons, has quasi-averages
\begin{equation}
<b_{\pm Q}> = <b_{\pm Q}^{\dagger}> = \sqrt{N_0 /2},
\label{eq:genBogolubov}
\end{equation}
corresponding to a condensate wave-function $\Psi_Q(y) \propto cos(Q
y)$. More generally $\Psi_Q(y)$ is multiplied by a phase factor
$e^{i \phi({\bf r},t)/\hbar}$, which is crucial to the BEC dynamics,
but not necessary for a stability analysis of the BEC.

We make a Holstein-Primakoff magnon expansion \cite{HP40} up to
4th-order in magnon operators, including contributions from both (2
in - 2 out) and (3 in - 1 out) magnon scattering processes \cite{sparks61}
(we ignore multiple-scattering contributions here, which are $\sim O(U_d/J_o)
\sim 10^{-3}$ relative to the leading terms). Then, taking quasi-averages, we
can write the Hamiltonian in the form
\begin{equation}
{\cal H} = \hbar \sum_q \omega_{\bf q} (b_{\bf q}^{\dagger} b_{\bf
q}+{1 \over 2}) + \widehat{V}^{p}_{int} + \widehat{V}^{-p}_{int}
\label{H-int}
\end{equation}
where the interaction term $\widehat{V}^{p}_{int}$ takes the form
\begin{eqnarray}
\widehat{V}^{p}_{int} = n_{0} (\Gamma_0 + {\Gamma_S \over 4})
\sum_{p<Q} \big[ b_{Q+p}^{\dagger}b_{Q+p} +
b_{-Q-p}^{\dagger}b_{-Q-p} \big] \;
\nonumber \\
+\frac{\Gamma_S n_0}{4} \sum_{p<Q} \big[ b_{Q+p}^{\dagger}b_{-Q+p} +
b_{-Q+p}^{\dagger}b_{Q+p} + \;\;
\nonumber \\
+ b_{Q+p}b_{-Q-p} + b_{Q+p}^{\dagger}b^{\dagger}_{-Q-p} \big] \; \nonumber \\
+ \frac{\Gamma_0 n_0}{2} \sum_{p<Q} \big[ b_{Q+p}b_{Q-p} +
b_{-Q-p}b_{-Q+p} + \;\;
\nonumber \\
+ b_{Q+p}^{\dagger}b^{\dagger}_{Q-p} +
b_{-Q-p}^{\dagger}b_{-Q+p}^{\dagger} \big].
   \label{Vint}
\end{eqnarray}
Here $\Gamma_0$ and $\Gamma_S$ are the four-magnon scattering
amplitudes between states $({\bf Q}, {\bf Q}) \rightarrow ({\bf Q},
{\bf Q})$ and $({\bf Q}, {\bf -Q}) \rightarrow ({\bf Q}, {\bf -Q})$
respectively. For the sample geometry in Fig.\ref{Fig1}, with ${\bf
H}_o \parallel \hat{x}$, these scattering amplitudes are found to be
\begin{eqnarray}
\Gamma_0 = - \frac{\hbar \omega_M}{8 S} \big[ (\alpha_1 - \alpha_3) {\cal F}_Q
- 2 \alpha_2 (1 - {\cal F}_{2Q}) \big] \;\;\;\;\;\;\;\;
\nonumber \\
- \frac{J Q^2}{4 S} \big[ \alpha_1 - 4 \alpha_2 \big];
\label{Gamma_0} \\
\Gamma_S = \frac{\hbar \omega_M}{2 S} \big[ (\alpha_1 - \alpha_2) (1 - {\cal F}_{2Q})
- (\alpha_1 - \alpha_3) {\cal F}_Q \big] \; \nonumber \\
+ \frac{J Q^2}{S} \big[ \alpha_1 - 2 \alpha_2 \big],
\label{Gamma_S}
\end{eqnarray}
with $\alpha_1 = u^4_Q + 4 u^2_Q v^2_Q + v^4_Q$, $\alpha_2 = 2 u^2_Q
v^2_Q$ and $\alpha_3 = 3 u_Q v_Q (u^2_Q + v^2_Q)$, where $\{ u_q, v_q \} =
\big[ (A_q \pm \hbar \omega_q) / 2 \hbar \omega_q \big]^{1/2}$ and
$A_q$ and $B_q$ are given by $A_q = \gamma H_i + J q^2 + 2 \pi \gamma M_s
{\cal F}_q$ and $B_q = - \pi \gamma M_s {\cal F}_q$ respectively. Higher-order
multiple-scattering contributions to $\Gamma_0, \Gamma_S$ are $\sim O(U_d/J_o)$
relative to the leading terms given here.

\begin{figure}[h]
\vspace{-1.8cm}
\includegraphics[width=8cm]{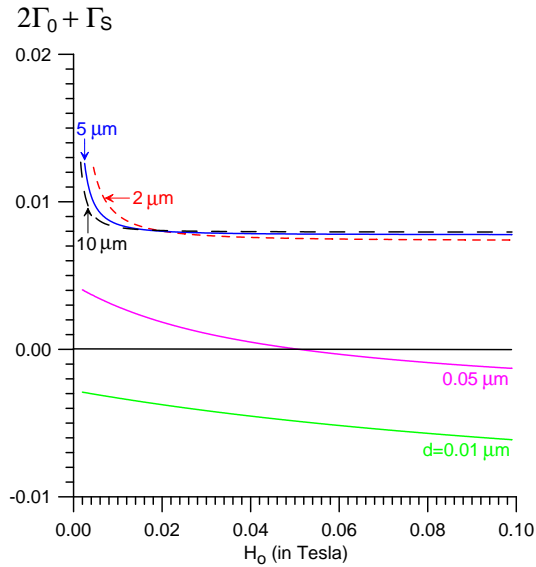}
\vspace{-1.8cm} \caption{ (color online). The effective amplitude $2
\Gamma_0 + \Gamma_S$ (in Kelvins) as a function of field $H_o$ at
different values of the film thickness $d$. Dashed lines: $d = 2$
and $10 \; \mu m$. Solid lines: $d = 0.01, \; 0.05$ and $5 \; \mu
m$.}
\label{Fig2}
\end{figure}

\begin{figure}[h]
\vspace{-1.9cm}
\includegraphics[width=8cm]{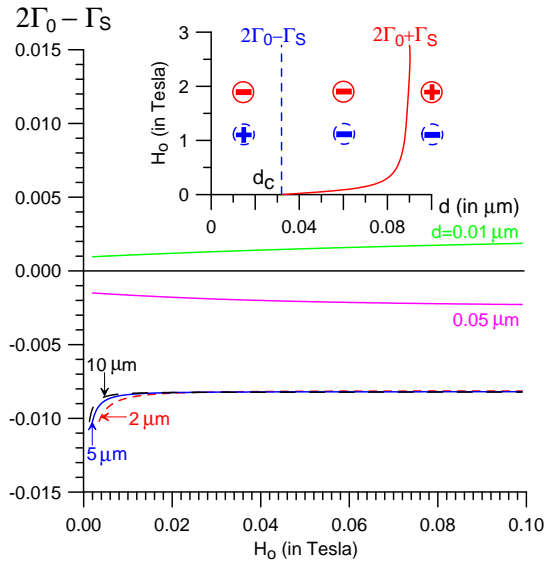}
\vspace{-1.7cm} \caption{ (color online). The effective amplitude $2
\Gamma_0 - \Gamma_S$ (in Kelvins) as a function of external field
$H_o$ at different values of the film thickness $d$. Dashed lines:
$d = 2$ and $10 \; \mu m$. Solid line: $d = 0.01, \; 0.05 $ and $5
\; \mu m$. The inset shows the "phase diagram" of YIG in the $(H_o,
d)$ plane. The amplitude $2 \Gamma_0 + \Gamma_S$ is positive
everywhere to the right of the solid line (the region $\oplus$). The
amplitude $2 \Gamma_0 - \Gamma_S$ is negative everywhere to the
right of the dashed line (the region $\ominus$). }
\label{Fig3}
\end{figure}

We plot the combinations $2 \Gamma_0 \pm \Gamma_S$ in Figures
\ref{Fig2} and \ref{Fig3}. Both these amplitudes are sensitive to
the external field and the film thickness. The first amplitude, $2
\Gamma_0 + \Gamma_S$, becomes negative in the entire region of
fields at $d < d_c \approx 2 (J / \hbar \omega_M)^{1/2} \approx
0.032 \; \mu m$. The second amplitude is positive at $d < d_c$.

Near the energy minimum (when $|p| << |Q|$), one has $\omega_{Q+p}
\approx \omega_{Q-p}$, and the Bogoliuibov transformation is
straightforward because ${\cal H}$ is symmetric when $p
\leftrightarrow -p$ and $Q \leftrightarrow -Q$. The spectrum thus
has 4 branches, with excited quasiparticle energies
\begin{eqnarray}
\epsilon_{p \eta} = \sqrt{\Omega_{Q}(p)[\Omega_{Q}(p) +n_{o}(2
\Gamma_{0}+ \eta \Gamma_{S})]}
\label{eq:spectr_symm}
\end{eqnarray}
where $\eta = \pm 1$, and $\Omega_{Q}(p) =\hbar(\omega_{Q+p} -
\omega_{Q})$.

The inset in Fig.\ref{Fig3} shows the "phase diagram" of the
quasi-two dimensional YIG for different values of $d$ and $H_o^x$.
One immediately sees a paradox: the amplitudes are never both
positive, so bulk BEC should not exist - yet BEC has been observed
\cite{demo06} in samples with an in-plane field $H_o^x \sim 0.07 \;
T$, in which $d \sim 5 \; \mu m$.

The paradox is resolved by noting that in a finite geometry, the
energy gap from the condensate to excited modes can be larger than
the scattering amplitude, leading to a potential barrier to decay.
For weakly-interacting Bose gases this yields \cite{BECmeta} an
upper critical number $n_o^{cr}$ of condensate particles, above
which the barrier disappears; one has $\alpha_o n_o^{cr}/l_o = k$,
where $\alpha_o$ is the s-wave scattering length, and $l_o$ the
characteristic size of the BEC wave function. The constant $k \sim
O(1)$, and depends on the sample geometry.

In the present case we can write the critical density $n_o^{cr}$  as
$|2 \Gamma_{0} - \Gamma_{S}| n_o^{cr} \sim  \epsilon_{\bf p}^{min}$,
where $ \epsilon_{\bf p}^{min}$ is the minimum quasiparticle energy
in the presence of the BEC; below this critical density the BEC is
metastable to tunneling or thermal activation. If the BEC were to
spread through the entire slab, then $n_o^{cr} |2 \Gamma_{0} -
\Gamma_{S}| = \Omega_Q(\pi/L)$, where the length $L$ depends on the
direction of the field relative to the slab axes. Taking this result
literally for the experiment \cite{demo06}, with a slab measuring
$20 \times 2 \; mm^2$ in the plane, one finds $10^{-8} < n_o^{cr} <
10^{-6}$ (for fields along the long and short sides of the slab
respectively). However this result is certainly too low, since it
assumes a perfectly uniform BEC - in reality disorder and edge
effects will smear the magnon spectrum and restrict the size of the
BEC, and a more realistic estimate for $L$ is then $L_{eff} =
(L_xL_yL_z)^{1/3} \sim 0.6 \; mm$. This yields $n_o^{cr} \sim
10^{-5}$ for the experiment. Thus we conclude that for in-plane
fields, it will be impossible to raise $n_o$ above this value; this
could be checked experimentally (eg., by increasing the pumping
rate).

\vspace{1mm}

\begin{figure}[h]
\vspace{-2.9cm}
\includegraphics[width=8cm]{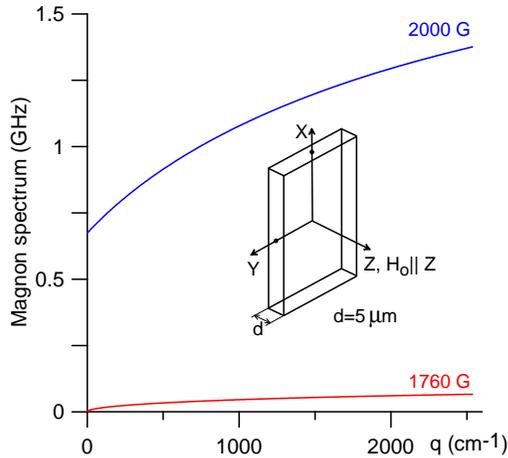}
\vspace{-2.9cm} \caption{ (color online). The magnon spectrum for an
infinite slab of YIG, assuming free surface spin boundary
conditions, with magnetisation polarised perpendicular to the plane
(see inset). The spectra are shown for $H_o^z = H_c \approx 1760 \;
G$ where the spectrum is gapless, and for $H_o^z = 2000 \; G$. }
\label{Fig4}
\end{figure}

{\bf (ii) Perpendicular field}: Now the results are very different.
Consider first an infinite thin slab, which is simple to analyze.
The competition between the external field ${\bf H_o} = \hat{z} H_o$
and the demagnetization field (which favours in-plane magnetisation)
gradually pulls the spins out of the plane; below a critical field
$H_c = \hbar \omega_M / \gamma \approx 1760 \; G$, the Free Energy
is degenerate with respect to rotation around $\hat{z}$ and so the
magnons are gapless, but at $H_c$ they align with ${\bf H_o}$ and a
gap $\hbar \omega_o \equiv \hbar \omega_{q = 0} = \gamma H_o - \hbar
\omega_M$ opens up. The minimum in the magnon spectrum is always at
$q=0$ (Fig.\ref{Fig4}).

The inter-magnon scattering amplitude $\Gamma$ is now always
positive; neglecting a very small exchange contribution one finds
\begin{eqnarray}
\Gamma({\bf q}) &=& (\hbar \omega_M  / 4 S) [1 - (1 - {\cal F}_q)/2]
\nonumber \\
&\rightarrow& (\hbar \omega_M  / 4 S) [1 - q_{||} d / 4 \; + \;
O(q_{\parallel}^2)]
\label{gamma+}
\end{eqnarray}
where $q_{\parallel}$ is the momentum in the $xy$ plane, and ${\cal
F}_q$ takes the form  (\ref{F_oo}) but with $q_x \rightarrow
q_{\parallel}$. This radically changes the situation - now a BEC is
stable with no restriction on the condensate density. There are
however restrictions on $H_o$; when $2200 \; G < H_o^z < 3500 \; G$
the system has a ``kinetic instability'' \cite{Melkov}, in which the
pumping of the magnons at one frequency destabilizes the magnon
distribution, along with strong microwave emission.


In a real finite sample things are much more complicated. Even
without surface anisotropy the spins near the slab edge are put out
of alignment with the bulk spins by edge demagnetisation fields; and
surface anisotropy does the same to spins on the slab faces. However
in the central region of the sample, at distances further from the
surface than the exchange length, the spectrum returns to the
infinite plane form. For fields well above $H_c$, eg., for $H_o^z
\sim 2000 \; G$, all of the spins will be aligned along $\hat{z}$,
and (\ref{gamma+}) will then be valid everywhere.

In this case we have the striking result that a BEC of pumped
magnons should be possible with densities much higher than present.
To give an upper bound is complicated since the problem then becomes
essentially non-perturbative (similar to, eg., liquid $^4He$),
beyond the range of higher-order magnon expansions. However there
appears to be no obstacle in principle to raising $n_o/n \sim O(1)$.
At present the highest achievable density is probably limited by
experimental pumping strengths rather than any fundamental
restrictions. Such a high-density BEC existing at room temperature
would be extremely interesting, and certainly possess unusual
magnetic properties.

\vspace{1mm}

{\bf Remarks}: The 2 cases studied above are actually limiting cases
of a more general situation in which one can manipulate the
inter-magnon interactions by varying the field direction and
strength, and vary the upper critical density for BEC formation by
changing the sample geometry. Thus the analysis can be easily
generalized to long 'magnetic wires' or whiskers, and we also expect
that BEC will be stabilized there when the external field is
perpendicular to the sample axis, but unstable or metastable when
the field is parallel to the sample axis. Further details of the
various possible cases will be published elsewhere.

\vspace{1mm}

{\bf Acknowledgements}: We acknowledge support by the Pacific
Institute of Theoretical Physics, the National Science and
Engineering Council of Canada, the Canadian Institute for Advanced
Research, the Louisiana Board of Regents (Contract No. LEQSF
(2005-08)-RD-A-29), the Tulane University Research and Enhancement
Fund, and the US Air Force Office of Scientific Research (Award no.
FA 9550-06-1-0110). We would also like to thank B. Heinrich and D.
Uskov for very useful conversations.

\end{document}